\documentclass[aps,prl,twocolumn,floatfix]{revtex4}
\usepackage{graphicx}
\usepackage{dcolumn}
\usepackage{bm}
\usepackage{amsmath}
\usepackage{epstopdf}
\usepackage{hyperref}

\usepackage{epstopdf, epsfig}
\usepackage{textcomp}
\usepackage{color}

\newcommand{\be}{\begin{equation}}
\newcommand{\ee}{\end{equation}}

\usepackage{blindtext}
\usepackage{lettrine}

\begin{document}

\title{Anomalous vortex motion induced by asymmetric vorticity distribution in rapidly rotating thermal convection}

\author{Shan-Shan Ding$^{1}$}
\author{Kai Leong Chong$^{2}$}
\author{Jun-Qiang Shi$^{1}$}
\author{Guang-Yu Ding$^{2}$}
\author{Hao-Yuan Lu$^{1}$}
\author{Ke-Qing Xia$^{3,2}$}
\email{email:kxia@cuhk.edu.hk}
\author{Jin-Qiang Zhong$^{1}$}
\email{email:jinqiang@tongji.edu.cn}

\affiliation{$^{1}$School of Physics Science and Engineering, Tongji University, Shanghai 200092, China \\
$^{2}$Department of Physics, The Chinese University of Hong Kong, Shatin, Hong Kong, China\\
$^{3}$Center for Complex Flows and Soft Matter Research and Department of Mechanics and Aerospace Engineering, Southern University of Science and Technology, Shenzhen 518055, China}

\date{\today}

\begin{abstract}
In rotating Rayleigh-B\'enard convection, columnar vortices advect horizontally in a stochastic manner. When the centrifugal buoyancy is present the vortices exhibit radial motions that can be explained through a Langevin-type stochastic model. Surprisingly, anomalous outward motion of cyclones is observed in a centrifugation-dominant flow regime, which is contrary to the well-known centrifugal effect. We interpret this phenomenon as a symmetry-breaking of both the population and vorticity magnitude of the vortices brought about by the centrifugal buoyancy. Consequently, the cyclones submit to the collective vortex motion dominated by the strong anticyclones.  Our study provides new understanding of vortex motions that are widely present in many natural systems.    
\end{abstract}

\maketitle

The emergence of coherent vortex structures is a prominent feature of rotating turbulent flows. The dynamics of the vortex structures plays an important role in determining fluid motions and turbulent transport, ranging from small-scale turbulence to large-scale geophysical flows \cite{Va06, Mc06, HV93, FS01}. Previous studies of the vortex dynamics are mostly focused on isolated vortices \cite{GL81, VK89, HV93}. However, fluid flows with densely populated vortices that exhibit long-range interactions often arise in rapidly rotating turbulence \cite{FS01, AMetal18}. Understanding the motions of, and interactions between these vortices is challenging. The fluid dynamics of rotating, buoyancy-driven turbulent flows is often studied using a paradigmatic model, rotating Rayleigh-B\'enard convection (RBC), i.e., a fluid layer heated from below and rotating about a vertical axis. Although considerable progress has been achieved in exploring non-rotating RBC \cite{AGL09, LX10, Xi13}, relatively few studies have been devoted to understand the flow structures and dynamics in rotating RBC. Recent studies report that when the applied rotation rate $\Omega$ increases, transitions in flow regimes \cite{VE02, KCG08, WW08, KSNHA09, KA12} occur from buoyancy-dominated convection with global circulations formed by self-organized plumes, to rotation-dominant turbulence where the convective flows are organized into thin coherent columnar vortices \cite{BG86, BG90, ZES93, Sa97, GJWK10, SLJVCRKA14}. Although a theoretical description of the columnar vortex structure has been provided by the Taylor-Proudman theorem \cite{Pr16, Ta23} that predicts fluid motion to be invariant along the vertical axis, the dynamics of horizontal advection of these vortices remains to be better understood. How is the horizontal vortex motion influenced by the buoyancy forcing, the rotation strength and the hydrodynamic interactions between adjacent vortices? It is a natural question of fundamental interest.  

The columnar vortices are helical structures with their vertical vorticity correlated to the flow directions, such that near the bottom boundary cyclones are upwelling (warm) vortices while anticyclones are downwelling (cold) ones, respectively \cite{Sa97, KA12}. In rotating RBC with a Boussinesq working fluid, a symmetric state was assumed in which the vorticity magnitude and the motion of both types of vortices are similar if observed in the midplane of the fluid layer. In rapidly rotating convection, however, such a flow symmetry may be broken when the centrifugal force destroys the horizontal translation invariance of the flow field. A recent direct numerical simulation (DNS) represents a first step in studying how centrifugal force impacts the heat-transport and the flow morphologies in turbulent convection \cite{HA18}. However, many rich dynamics of vortex motions under the influence of centrifugal force remain unexplored.

In this Letter we present experimental and numerical studies that shed new light on the nature of vortex motions and their interactions in rapidly rotating convection. We first show that, as expected, vortices that undergo centrifugal buoyancy exhibit radial directional motion, i.e. warm cyclones (cold anticyclones) move inwardly (outwardly), and these motions are influenced by the turbulent background fluctuations. Our surprising finding is a new type of vortex motion in which cyclones move outwardly in a centrifugation-dominant regime. This phenomenon can be understood in terms of a centrifugal-buoyancy induced broken symmetry between the two types of vortices, in that both the number and the vorticity of the anticyclones become larger than the cyclonic ones. Remarkably, the cyclones submit to the collective mode of long-range correlated motion dominated by the anticyclones.     
  
\begin{figure}
\includegraphics[width=0.5\textwidth]{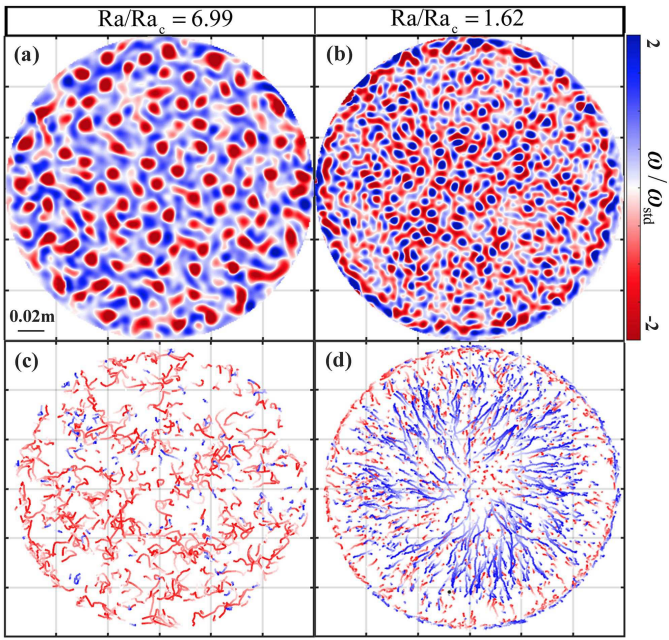}
\caption{(a, b) Experimental data of the instantaneous vertical vorticity distribution $\omega/{\omega}_{std}$ over the cross-section of the cell. ${\omega}_{std}$ is the standard deviation of $\omega$. (c, d) Trajectories for cyclones (red) and anticyclones (blue). The shading of the trajectories indicates that each vortex appears at the light-color side and terminates at the dark-color side. Results for $\mathrm{Ra}{=}3.0{\times}10^{7}$ and $\mathrm{Ra/Ra_c}{=}6.99$, Fr${=}0.03$ (a, c);  $\mathrm{Ra/Ra_c}{=}1.62$, Fr${=}0.27$ (b, d).}
\label{fig:1}
\end{figure} 

The convection apparatus was designed for high-precision heat transport and flow structure measurement in rotating RBC \cite{ZSL15, SLZ16, ZLW17}. To visualize the flows we used two cylindrical cells with sapphire top windows. Both had a circular copper bottom plate with a Plexiglas side wall with inner diameter $d{=}$240 mm. The long (short) cell had a fluid height $H{=}$120.0 (63.0) mm, yielding the aspect ratio $\Gamma{=}d/H{=}$2.0 (3.8). The experiment was conducted with a constant Prandtl number Pr${=}\nu/{\kappa}{=}4.38$ and in the range $2.0{\times}10^7{\le}$Ra${\le}2.7{\times}10^8$ of the Rayleigh number Ra${=}{\alpha}g{\Delta}TH^3/{\kappa}{\nu}$ ($\alpha$ is the isobaric thermal expansion coefficient, $g$ the acceleration of gravity, $\Delta T$ the applied temperature difference, $\kappa$ the thermal diffusivity, and $\nu$ the kinematic viscosity). Rotation rates up to 5.0 rad/s were used. Thus the Ekman number Ek${=}\nu/2{\Omega}H^2$ spanned $4.9{\times}10^{-6}{\le}$Ek${\le}2.7{\times}10^{-4}$, corresponding to a range of the reduced Rayleigh number $1.06{\le}\mathrm{Ra/Ra_c}{\le}120$, with $\mathrm{Ra_c}{=}8.7$Ek$^{-4/3}$ the critical value for the onset of convection \cite{Ch61}. The Froude number Fr${=}{\Omega}^2d/2g$ was within $0{<}$Fr${\le}0.31$. The flow field at a fluid depth of $z{=}H/4$ was measured using the technique of particle image velocimetry (PIV). We obtained horizontal velocity $\vec{v}(r,{\phi})$ over the full cross-section of the cell. In the DNS \cite{SM} we solved the Navier-Stokes equations with the Coriolis and centrifugal forces included, using the multiple-resolution version of the {\it{CUPS}} \cite{KX13, KSX14, CDX18}. The simulation was performed in a cylindrical cell with $\Gamma{=}4$, $\mathrm{Ra}{=}2.0{\times}10^{7}$ and $1.06{\le}\mathrm{Ra/Ra_c}{\le}40$.

\begin{figure}
\includegraphics[width=0.5\textwidth]{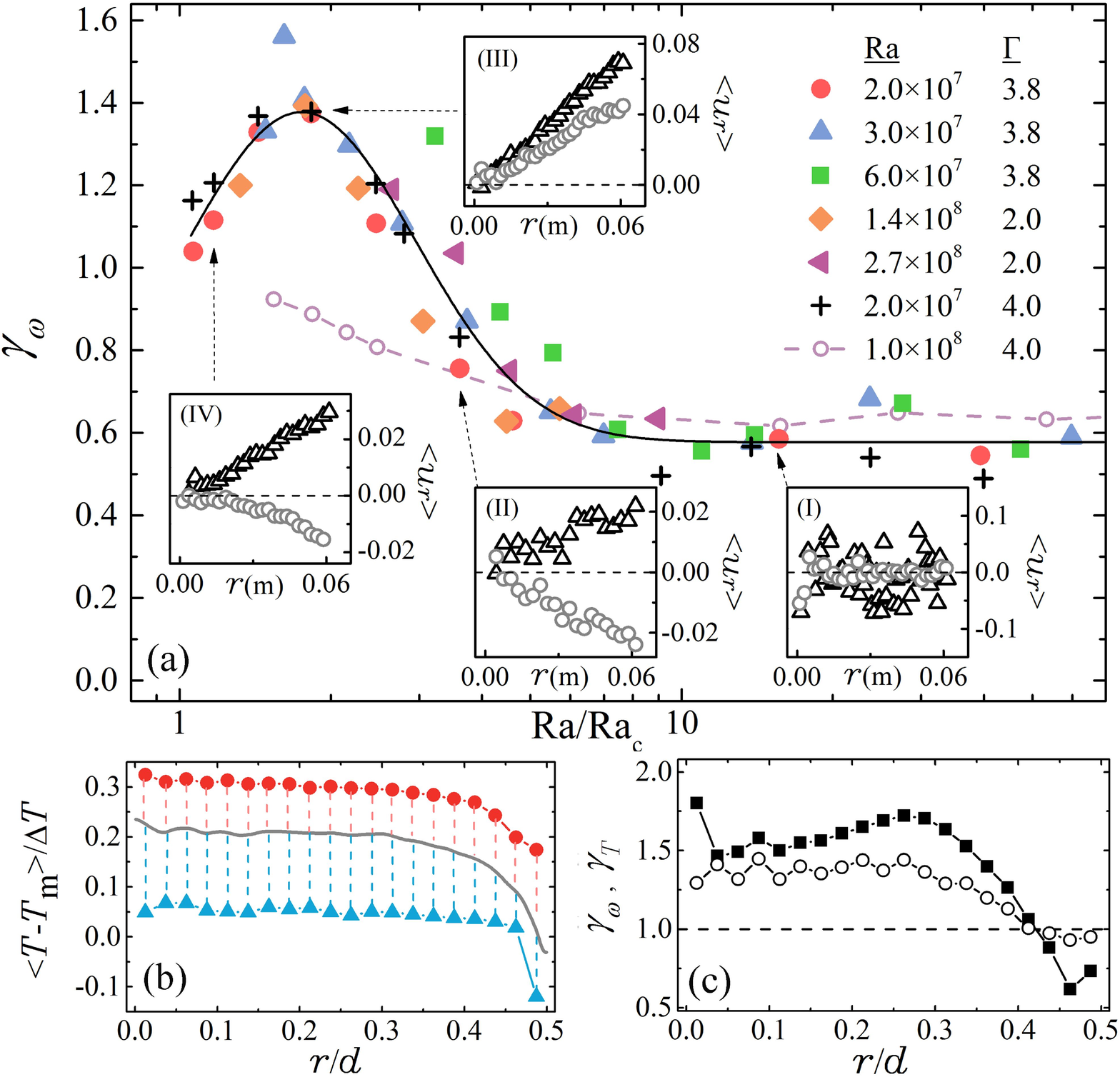}
\caption{(a) The vorticity ratio $\gamma_{\omega}{=}{\lvert}\langle{\omega}_{a}\rangle/\langle{\omega}_{c}\rangle{\rvert}$ as a function of $\mathrm{Ra/Ra_{c}}$. $\langle{\omega}_{a}\rangle$ and $\langle{\omega}_{c}\rangle$ denote the mean of ${\omega}$ measured at the centroids of the anticyclones and cyclones in the inner region. Filled symbols: experimental data. Pluses: data from DNS that includes the centrifugal force. Open circles: data from DNS that neglects the centrifugal force. DNS data are acquired at $z{=}0.2H$. The solid curve indicates the tendency of the experimental data. Inset panels: radial profiles $\langle{u}_r(r)\rangle$ (in unit of $mm/s$) of cyclones (circles) and anticyclones (triangles) for Ra${=}2.0{\times}10^7$ and from right to left, $\mathrm{Ra/Ra_{c}}{=}15.6, 3.61,1.83,1.17$. (b) Radial profiles of the mean temperature $\langle |T{-}T_m|\rangle/{\Delta T}$ for the background fluid (gray curve), cyclones (red circles) and anticyclones (blue triangles). $T_m$ is the arithmetic mean of the top and bottom fluid temperature. The length of the dashed lines indicates the temperature difference $\delta{T}$ between the cyclones (anticyclones) and the background fluid. (c) Radial profiles of ${\gamma}_{\omega}$ (open circles) and ${\gamma}_{T}{=}\lvert{\langle}{\delta T_{a}}{\rangle}/{\langle}{\delta T_{c}}{\rangle}\rvert$ (solid squares). ${\langle}{\delta T_{a}}{\rangle}$ and ${\langle}{\delta T_{c}}{\rangle}$ denote the mean of $\delta{T}$ of anticyclones and cyclones. (b) and (c) are DNS data for Ra${=}2.0{\times}10^7$, $\mathrm{Ra/Ra_{c}}{=}1.83$.}
\label{fig:2}
\end{figure} 

Figure 1 presents snapshots of the vortex structures. At a low rotation rate (Fig.\ 1a), it appears that over the measured fluid height cyclonic vortices (shown as red vorticity patches) possess a greater number density and on average larger vorticity than that of anticyclones (blue patches). Both types of vortices exhibit stochastic horizontal motions as indicated by the vortex trajectories in Fig.\ 1c, presumably driven by the turbulent background flows. The mean-square-displacement (MSD) of the vortices approaches to a liner function of time at large times, indicating a Brownian-type, normal diffusive motion \cite{Paper1}. At higher rotation rates, however, we observe strong anticyclones with larger population (Fig.\ 1b). The anticyclones undergo outward radial motions accompanied by stochastic fluctuations along their paths (Fig.\ 1d). Part of the anticyclonic motions are terminated by the retrogradely traveling plumes near the sidewall. On the other hand, the weak cyclones display complex motions. Figure 1d indicates that in the outer region ($r{\ge}d/4$) the cyclones move toward the cell center while in the inner region ($r{\le}d/4$) most of them migrate radially outwardly. We note that the MSD of both types of vortices indicate superdiffusive behavior in this flow regime (see Supplementary Movies \cite{SM} of the vortex motions). 

To further analyze the complex vortex motions, we show in the insets of Fig.\ 2a profiles of the mean radial velocity  ${\langle}u_r{\rangle}$ of the vortices measured in the inner region of the cell. These velocity profiles illustrate that depending on the rotation rates there are four distinct flow regimes. A randomly-diffusive regime exists in the slow rotating limit with Ra one order in magnitude larger than $\mathrm{Ra_c}$. In this regime the vortices move in a random manner with ${\langle}u_r{\rangle}{\approx}0$ for both cyclones and anticyclones (Panel (I)). When $\Omega$ increases, the magnitude of ${\langle}u_r{\rangle}$ increases linearly with $r$ in a centrifugally-influenced regime ($3\mathrm{Ra_c}{\le}\mathrm{Ra}{\le}5\mathrm{Ra_c}$), where we observe that cyclone moves inward radially, in opposite to the outward anticyclonic motion. In a centrifugation-dominant flow regime with $1.5\mathrm{Ra_c}{\le}\mathrm{Ra}{\le}3\mathrm{Ra_c}$, anomalous \emph{outward cyclonic motion} is observed (Panel (III)). In this anomalous regime the radial gradient of ${\langle}u_r{\rangle}$ for both types of vortices reach a maximum. In the rapid rotation limit $\mathrm{Ra}{\le}1.5\mathrm{Ra_c}$ (asymptotic regime), ${\langle}u_r{\rangle}$ decreases and the opposite radial motions of cyclones and anticyclones recover.     

Also plotted in Fig.\ 2a is the vorticity ratio $\gamma_{\omega}$ of the anticyclones to the cyclones. In the randomly-diffusive regime $\gamma_{\omega}$ is near 0.6, indicating that cyclones have larger vorticity in magnitude than anticyclones, which is also evident in Fig.\ 1a. This asymmetry in vorticity strength is owing to the fact that upwelling cyclones penetrate the fluid layer at $z{=}H/4$ (where PIV measurement is made). However, downwelling anticyclones generated from the top hardly reach this position, as their momentum and vorticity are partially dissolved by the background turbulence \cite{KCG10}. With increasing $\Omega$ the up- and downwelling vortices evolve and in the rapidly rotating limit they form vertically symmetric, columnar structure under the TP constraint \cite{GJWK10, SLJVCRKA14}. Hence one would expect that the ratio $\gamma_{\omega}$ increases and approaches unity monotonously. This is the case if the centrifugal effect is absent, as verified by our DNS data (Fig.\ 2a). The experimental data show, however, that in the centrifugally-influenced regime $\gamma_{\omega}$ increases rapidly, and exceeds unity in the anomalous regime, indicating an anticyclone-dominated flow field as shown in Fig.\ 1b. In the asymptotic regime where the severe rotational constraint finally weakens the convective vortices, the data indicate that $\gamma_{\omega}$ approaches unity and the symmetry between cyclonic and anticyclonic vorticity restores. It is worth noting that various data sets of $\gamma_{\omega}(\mathrm{Ra}/\mathrm{Ra_c})$ with Ra spanning two decades and with different aspect ratios collapse well onto a universal curve. 

\begin{figure}
\includegraphics[width=0.5\textwidth]{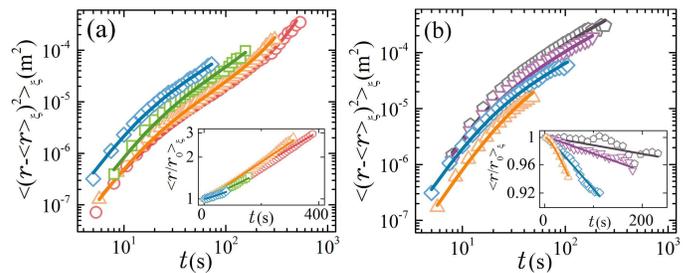}
\caption{Experimental data (open symbols) and theoretical predictions (solid lines) of ${\langle}(r{-}{\langle}r{\rangle}_{\xi})^{2}{\rangle}_{\xi}$ for anticyclones (a) and cyclones (b). Results for Ra${=}3.0\times 10^7$ and $\mathrm{Ra}/\mathrm{Ra_c}{=}1.62$ (circle), 1.77 (up triangle), 2.77 (square), 3.74 (diamond), 6.99 (down triangle), 13.81 (pentagon). The insets show the results of ${\langle}r/r_0{\rangle}_{\xi}$. ${\langle}...{\rangle}_{\xi}$ denotes a trajectory-assemble mean. $r_0$ is the initial radial position of a vortex.}
\label{fig:3}
\end{figure} 

The centrifugal effect can be shown to be responsible for the observed radial motion of the vortices. We consider a model that consists of a set of Langevin equations:
\begin{equation}
\ddot r+\dot r/\tau \pm \zeta r=\xi(t).
\label{eq:one}
\end{equation}
The radial acceleration of the vortices $\ddot{r}$ is determined by the viscous damping ${\dot{r}/\tau}$ and the centrifugal force ${\pm}{\zeta}r$ that drives cold (warm) fluid radially away from (towards) the rotation axis. The background turbulent fluctuations as well as the stochastic vortex-vortex interactions are modeled by a $\delta$-correlated, Gaussian white noise term $\xi(t)$ with ${\langle}\xi(t)\xi(t{+}{\Delta}t){\rangle}{=}{\delta}({\Delta}t)D/\tau^2$ \cite{SBN02, BA07}. Here $\tau$ is the characteristic timescale of the ballistic regime constrained by viscous damping; $D$ is the diffusivity that reveals the strength of the background fluctuations; and $\zeta$ is the centrifugal coefficient \cite{SM}. In the slow rotating limit, $\zeta$ approaches zero and the model reduces to the classical Brownian-motion model \cite{La08}. 

We compute the first and second moments of the radial vortex displacements according to Eq.\ 1 \cite{SM}, and compare in Figs.\ 3a and 3b these theoretical predictions with the experimental data (here we have excluded the data of outward cyclonic motion in the anomalous regime). The first moments of the radial displacement ${\langle}r/r_0{\rangle}_{\xi}$ are sums of multiple exponential functions of time, and approach a single exponential function as $exp({\pm}{\lambda^{\ast}}t)$ in the large-time limit, with the growth rate ${\lambda}^{\ast}{=}|1/{2\tau}{-}\sqrt{1/{4\tau ^2}{\pm}\zeta}|$ that reveals the mobility of the vortex motion. The second moments of the radial displacement, ${\langle}(r{-}{\langle}r{\rangle}_{\xi})^{2}\rangle_{\xi}$, represent the mean deviations of the vortex trajectories from their mean path, or the ``track forecast cones" \cite{MPSBK12} of the vortices. Notably the theoretical curves of ${\langle}(r{-}{\langle}r{\rangle}_{\xi})^{2}\rangle_{\xi}$ appear differently for cyclones and anticyclones. In a log-log plot, the second moment of anticyclonic displacement takes on the feature of superdiffusion at large times, as it increases steeply after an inflection point, whereas the curves of the cyclonic data appear to level off. These predicted trends agree with the experimental data over two decades in time and three orders of magnitude in the displacements. Detailed parametrization ($\tau, D, {\lambda^{\ast}}$) of the model is given in the Supplemental Material \cite{SM}. 

The close agreements between the model and the experimental data shown in Fig.\ 3 suggest that the simple stochastic model (Eq.\ 1), which incorporates the centrifugal effect, captures the basic physics that governs the main dynamics of the vortex motion. Nonetheless, the outward cyclonic motion observed in the anomalous regime (Figs.\ 1d and 2a) are unexpected. In this centrifugation- dominant flow regime, we observed the maximum vorticity ratio $\gamma_{\omega}$ and radial gradient of ${\langle}u_r{\rangle}$ for both types of vortices (Fig.\ 2a). Thus we probe further into the centrifugal effects on the vortex structures. Figure 2b shows the radial profiles of the mean temperature of the background fluid and of the vortices. These numerical data indicate a noticeable warming of the background fluid in the inner region, which is due to the accumulated warmer fluid, a centrifugal effect that has been observed \cite{HO99, LE11}. As a result the temperature difference ${\delta}T$ between the cold anticyclones and the background fluid exceeds that between the warm cyclones and the background (Fig.\ 2b). Since ${\delta}T$ is proportional to the buoyancy forcing on the vortex core, it is predicted to be positively correlated to the vorticity $\omega$ in recent theoretical models \cite{PKVM08, GJWK10}. Figure 2c shows that ${\delta}T$ and $\omega$ exhibit a similar trend and in particular both are larger for anticyclones than for cyclones in the inner region, which also explains the results of the vorticity ratio in the anomalous regime.

 \begin{figure}
\includegraphics[width=0.5\textwidth]{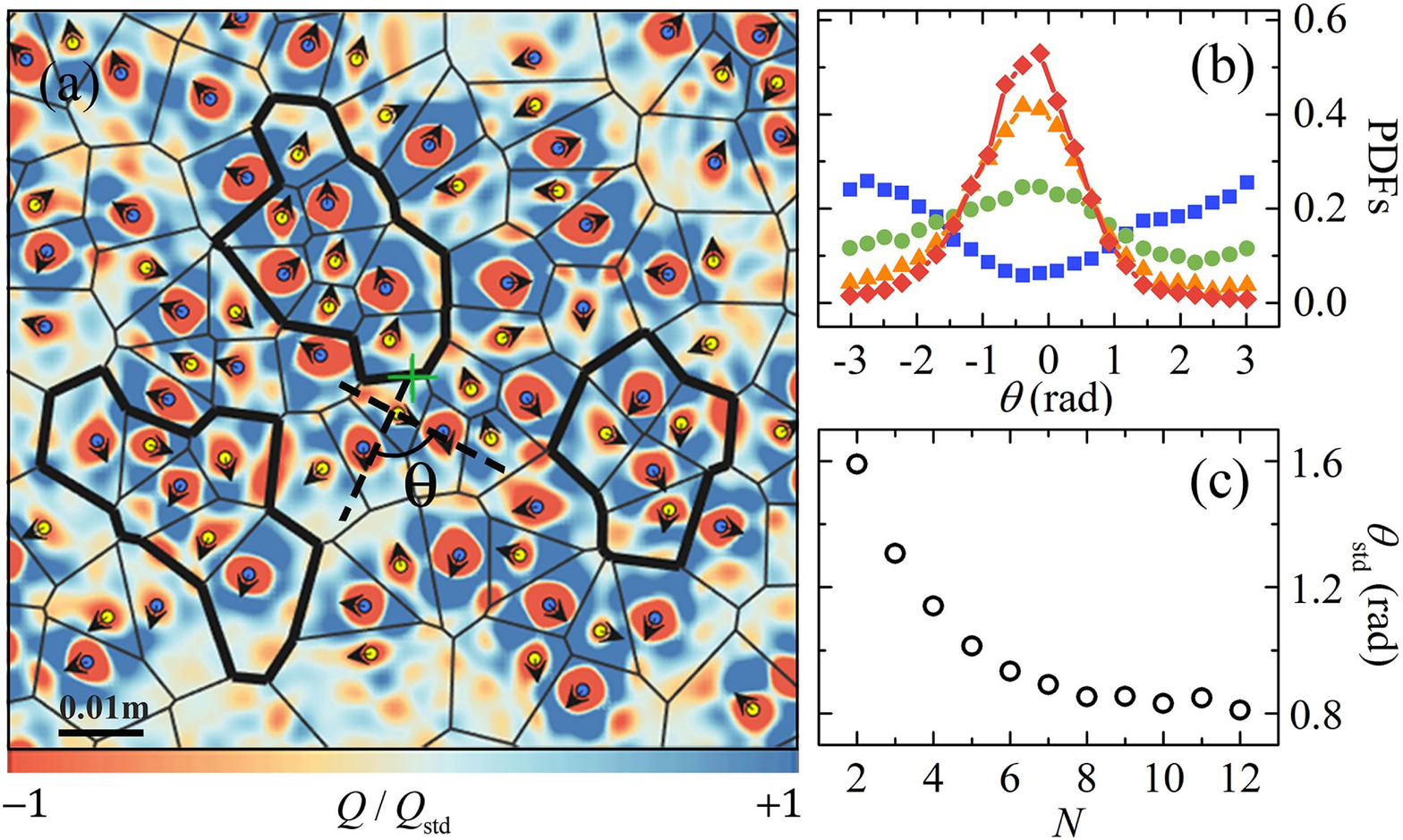}
\caption{(a) Snapshot of the vortex distribution and instantaneous velocity of the vortices in the central square region for $\mathrm{Ra}{=}3.0{\times}10^7$, $\mathrm{Ra}/\mathrm{Ra_c}{=}1.62$.  Blue (yellow) circles denote the centers of anticyclones (cyclones). Black arrows denote the velocity direction of the vortices. The solid-line network represents the Voronoi diagram of the vortex centers. Examples of three vortex clusters are marked with thick boundaries. The background coloration represents the distribution of the quantity $Q/Q_{std}$ \cite{SM}. $\theta$ is the angle between the position vector of cyclones relative to the cell center (the green cross) and their velocity. (b) Probability density functions of $\theta$ for isolated cyclones ($N{=}1$, blue square) and clustered cyclones (green circles: $N{=}2$; orange triangles: $N{=}4$; red diamonds: $N{=}7$). (c) Standard deviation of $\theta$ as a function of $N$.}
\label{fig:4}
\end{figure} 
    
To understand the anomalous motion of cyclones we examine the collective motion of the vortices that occurs noticeably in the anomalous regime. Figure 4a is an example of the instantaneous motion of the vortices, with their spatial distribution presented in a Voronoi diagram. We see that adjacent vortices often self-organize into vortex clusters, in the sense that vortices within each cluster largely move in the same direction. For quantitative analysis, we adopt two criteria to identify vortex clusters, i.e., the distance of two neighboring vortices is smaller than 1.5 times the mean vortex diameter and the angle $\phi$ between their velocity vectors is within a threshold ($\phi{\le}\phi^{\ast}{=}60^{\circ}$). Here we did the analysis over the range $30^{\circ}{\le}\phi^{\ast}{\le}75^{\circ}$ to confirm that the results of correlated vortex motion are not sensitive to the choice of $\phi^{\ast}$. The direction of the radial motion of cyclones is represented by the angle $\theta$ between their position vector $\vec{r}$ relative to the cell center and their velocity $\vec{u}$. It is clear that $\theta$ is related to the number ($N$) of vortices in a cluster (Fig.\ 4b). For isolated cyclones ($N{=}1$), the most probable direction of motion is radially inward ($\theta_p{=}\pi$). When $N{>}1$ and cyclones move collectively with neighboring vortices, we find $\theta_p{=}0$ as cyclones move anomalously outward. The standard deviation of $\theta$ decreases monotonically when $N$ increases (Fig.\ 4c). On the basis of these analyses, we suggest that it is the emergence of the centrifugal effects that breaks the symmetry of the vortex distribution, leading to the larger magnitude of vorticity for anticyclones than for cyclones in the inner region. Meanwhile adjacent vortices are often organized into vortex clusters by means of hydrodynamic interaction \cite{JLMW96}. Within large clusters the motion of cyclones submit to that of the anticyclones that are subject to stronger centrifugal buoyancy. As a result, we observe the outward migration of cyclones in a collective manner, and their motions become more unidirectional with the increasing size ($N$) of the cluster.

We have shown in the example of densely populated vortices in rotating RBC that local hydrodynamic interaction
can lead to the emergence of collective vortex motion. As a result, weak cyclones are dynamically arrested and submit to the long-range correlated vortex motion dominated by the strong anticyclones. Our study sheds new insight on the collective motion in other systems, such as clustering and jamming of densely assembled particles in granular media \cite{LN10, Na17} and active matters \cite{Be14, BLLRVV16}, and thus has broad implications to the studies of soft-condensed-matter and fluid physics. 

This work is supported by the National Science Foundation of China under Grant No. 11572230 and 11772235, a NSFC/RGC Joint Research Grant No. 1561161004 (JQZ) and N\_CUHK437$/$15 (KQX) and by the Hong Kong Research Grants Council under Grant No. 14301115 and 14302317.

\end{document}